# Thermodynamic evidence for pressure-induced bulk superconductivity in the Fe-As pnictide superconductor CaFe$_2$As$_2$

Y. Zheng[1], Y. Wang[1], B. Lv[2], C. W. Chu[2], and R. Lortz[1]♣

[1]*Department of Physics, Hong Kong University of Science & Technology, Clear Water Bay, Kowloon, Hong Kong*
[2]*Department of Physics, Texas Center for Superconductivity, University of Houston (TCSUH), 202 Houston Science Center, Houston, TX 77204-5002, USA*

We report specific-heat and resistivity experiments performed in parallel in a Bridgman-type of pressure cell in order to investigate the nature of pressure-induced superconductivity in the iron pnictide compound CaFe$_2$As$_2$. The presence of a pronounced specific-heat anomaly at $T_c$ reveals a bulk nature of the superconducting state. The thermodynamic transition temperature differs dramatically from the onset of the resistive transition. Our data indicates that superconductivity occurs in the vicinity of a crystallographic phase transition. We discuss the discrepancy between the two methods as caused by strain-induced superconducting precursors formed above the bulk thermodynamic transition due to the vicinity of the structural instability.

With the recent discovery of the iron pnictide superconductors [1], the phenomenon of high-temperature superconductivity is no longer limited to cuprate materials. The opportunity of having two distinct high-temperature superconducting families represents great opportunities to gain new insights into the mechanism of high-temperature superconductivity. In contrast to the antiferromagnetic Mott insulators of the cuprates parent compounds, those of the pnictide superconductors show spin density wave (SDW) magnetism [2], although it is still debated whether the nature of magnetism is itinerant or localized [3-5]. Superconductivity is mainly associated with the Fe-As layers and appears close to the border of the antiferromagnetically ordered SDW phase [2,6]. A maximum $T_c$ of 55 K has been observed in the '1111' structural family of the electron-doped ReOFeAs compounds (Re: rare earth) [7]. Further classes of materials are the self-doped '111' compounds (AFeAs, where A = alkaline) and the hole-doped '122' compounds based on the MFe$_2$As$_2$ parent compound (M: Ba, Sr, Ca, K, Rb and Cs) with $T_c$ values of up to 38 K for K-doped BaFe$_2$As$_2$ and SrFe$_2$As$_2$ [8,9]. The 122 family is particular interesting for detailed studies of the pnictides phase diagram as large single crystals are available and the parent compound can be doped on either the M site [10-12], the Fe site (Co) [13] or the As site (P) [14] or by the application of pressure [15,16], not to mention the appearance of superconductivity in the undoped compounds under micrographic strain [17]. Doping charge carriers suppresses the magnetic SDW transition which more or less coincides with a structural transition from a low-temperature orthorhombic (O) to a tetragonal (T) phase. CaFe$_2$As$_2$ (Ca122) is the member with the smallest unit cell volume and moderate pressure in the kbar range is sufficient to induce superconductivity and a collapsed tetragonal (cT) phase in the parent, undoped compound [15,16,18,19]. Recently, relatively high superconducting transition temperatures have been reported in hole-doped (Ca$_{1-x}$Na$_x$)Fe$_2$As$_2$ (above 33 K) [9] and in electron-doped (Ca$_{1-x}$Pr$_x$)Fe$_2$As$_2$ Ca122 with $T_c$ values of up to 49 K [10,11]. This motivates further research to better understand the mechanism of

---

♣ Corresponding author: lortz@ust.hk

superconductivity and the exact phase diagram in the parent undoped compound Ca122 under pressure, especially as it has been suggested that a giant coupling of the on-site Fe-magnetic moment with the As-As bonding along the z-axis may provide a mechanism for unconventional superconductivity in the 122 materials [20].

Besides the O to T structural phase transition, which is suppressed at relatively low pressure, a further structural transition from the T to the cT phase occurs in higher pressures [19]. The transition is associated with a drastic reduction in the c/a ratio. Pressure-induced superconductivity in Ca122 is complicated by the fact that it occurs in the vicinity of the pressure-induced structural transition from the O to the cT phase. Theoretical investigations indicated that all structural transitions in Ca122 are strongly coupled to electronic degrees of freedom and are accompanied by a pronounced change of the Fermi surface [21,22]. Therefore, the presence of even small pressure gradients in the experiment may strongly influence the structural, electronic and superconducting properties [23]. The exact nature of pressure-induced superconductivity is hence strongly debated: In susceptibility and transport studies under hydrostatic pressure with helium as pressure medium an exceptionally sharp transition from the O phase to the non-magnetic cT phase has been found. At low temperatures only a very weak drop of resistivity has been observed without any feature in susceptibility, which suggests the absence of bulk superconductivity under purely hydrostatic conditions [24]. It has been proposed that, in order to observe pressure-induced superconductivity in Ca122, a certain uniaxial pressure component is required to stabilize a tetragonal phase at low temperatures which may coexist with other structural phases [25]. It has been furthermore argued that superconductivity under non-hydrostatic conditions may originate from the presence of a strain-induced multi-crystallographic mixed phase [19,26]. Superconductivity has been proposed to be associated with domain walls between highly phase-separated regions of different structures [24]. In this scenario, superconductivity would be of filamentary nature. AC-susceptibility measurements under pressure indicated however that, under quasi-hydrostatic conditions, a considerable volume fraction of the samples becomes superconducting [26].

In contrast to measurements of the electrical resistivity and the magnetic susceptibility, which are often fooled by filamentary or surface superconductivity [27], the specific heat is a true bulk thermodynamic method and perfectly suited to investigate whether pressure-induced superconductivity in Ca122 is of bulk nature. Furthermore, it is represents an ideal tool to investigate pressure-induced structural phase transitions. Measurements of the specific heat under pressure are difficult due to the need of thermally isolating the sample. In this paper, we present specific-heat data under pressure up to 20 kbar on a single crystal of Ca122 under quasi-hydrostatic conditions. As we will show, the observation of a pronounced specific-heat anomaly clearly indicates bulk superconductivity. The measurements furthermore provide details on the structural transitions upon approaching the superconducting phase and reveal a strong difference between the thermodynamic and the resistive determination of the superconducting transition temperature.

Single crystals of Ca122 have been grown from self-flux as described in detail elsewhere [11]. The specific-heat measurements have been performed in a Bridgman-type of pressure cell by the same method used before for cuprate superconductors [28] with steatite as pressure medium. The materials for the different components of the pressure cell have been chosen to carefully compensate the thermal expansion of the cell in order to ensure a constant pressure during temperature sweeps. Steatite (which is also called 'soap stone') is less hydrostatic than commonly used liquid pressure media such as pentane–isopentane or Daphne oil 7373 [29] but has nevertheless rather good quasi-hydrostatic conditions and the great advantage over most of the liquid media (especially the most hydrostatic medium helium), that it is easier to achieve isolating the sample thermally form its surroundings by an AC modulated-temperature specific-heat technique at fast modulation. The drawback is that, in presence of structural transitions which may be sensitive to non-hydrostatic pressure components, the phase diagram may be modified somewhat. On the other hand, as steatite is solid from the beginning, we avoid the additional stress induced upon cooling a liquid pressure medium through its solidification transition. Furthermore, it has been shown that under purely hydrostatic conditions Ca122 does not exhibit bulk superconductivity [24]. In order to investigate the pressure-induced superconductivity we therefore need to induce a certain non-hydrostaticity.

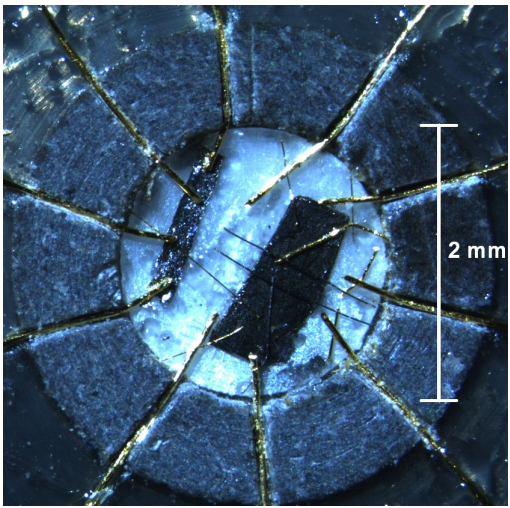

**Figure 1.** Photograph of the experiment mounted within the pyrophyllite gasket of a Bridgman type of pressure cell with a solid pressure transmitting medium, as taken through a microscope. The cell is shown before the second steatite disk had been placed into the gasket. The upper left thin strip represents the Pb manometer, the lower right black rectangle is the sample of Ca122.

In Figure 1 we present a photograph of the experiment mounted within a pyrophyllite ($Al_2Si_4O_{10}(OH)_2$) gasket (gray ring) on one of the tungsten carbide anvils of the Bridgman pressure cell. Ten grooves have been cut by hand into the gasket with help of a razorblade with embedded gold wires as electrical feedthroughs. The white background within the gasket is one disk of steatite on which the experiment has been mounted. The larger lower left black rectangle represents the Ca122 single crystal which has been polished from both sides down to a thickness of 20 µm. The thin gray strip on the upper left is a thin Pb foil in a 4-wire electrical configuration. The superconducting transition of lead is pressure dependent and, with help of literature data [30,31], it serves as a sensitive manometer. The sample is contacted with thin gold wires for measurement of the electrical resistivity in 4-wire configuration, as well as with one AuFe / Chromel and one Chromel / Constantan thermocouple for the heat-capacity measurements. After completion of the experimental setup a second disk has been put into the gasket to cover the experiment. The volume of the steatite is chosen to fill 2/3 of the volume within the gasket to reach optimal quasihydrostatic conditions. The gold wires in the grooves have been insulated electrically by pyrophillite powder which has been pushed into the grooves over the gold wires. The anvil has been put into the cylindrical body of the pressure cell and the gasket was squeezed between two anvils in order to apply the pressure.

The electrical resistivity was measured with a Keithley™ 6221 AC-current source in combination with a digital lock-in amplifier. The frequency was chosen as a few Hz in order to minimize phase shifts due to dissipation or capacitive effects. For the heat-capacity measurement we sent an AC current through one of the two thermocouples and used its contact resistant as a local Joule heater in order to periodically modulate the sample temperature. The second thermocouple was used as a sensor and both amplitude and phase shift of the thermocouple was measured with a lock-in amplifier. A standard model of AC calorimetry [34] relates the amplitude and phase shift of the induced temperature modulation $T_{AC}$ to the heat capacity ($C$) of the sample and the thermal conductance ($K$) between sample and the heat bath (which is represented by the anvils).

$$T_{AC} = P_0 / [K + i\omega C]$$

$P_0$ is the heating power provided to the sample. If the frequency $\omega$ of the temperature modulation is chosen sufficiently high (typically 200 Hz – 1 kHz), the heat capacity term dominates and the thermal conductance can be neglected in a good approximation. Note, that this model is too simple to exactly model the heat flow within the pressure cell. Furthermore, it is impossible to separate the heat capacity of the sample from that of the surrounding pressure transmitting material. For this reason, we do not attempt to extract absolute values of the specific heat but investigate relative changes in the heat capacity in the vicinity of thermodynamic phase transitions. Although the heat-capacity data is presented in arbitrary units, the data is normalized by the exact heating power provided and therefore allows us to compare the magnitude of the heat-capacity data taken under different pressures.

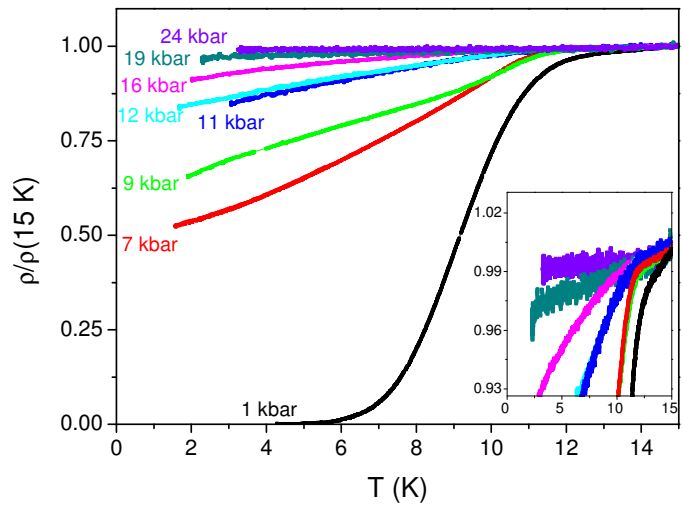

**Figure 2 a)** Electrical resistance of Ca122 for various applied pressures between 1 and 24 kbar. The data has been normalized at 15 K for clarity. The inset shows an enlargement on the onset of the superconducting transition.

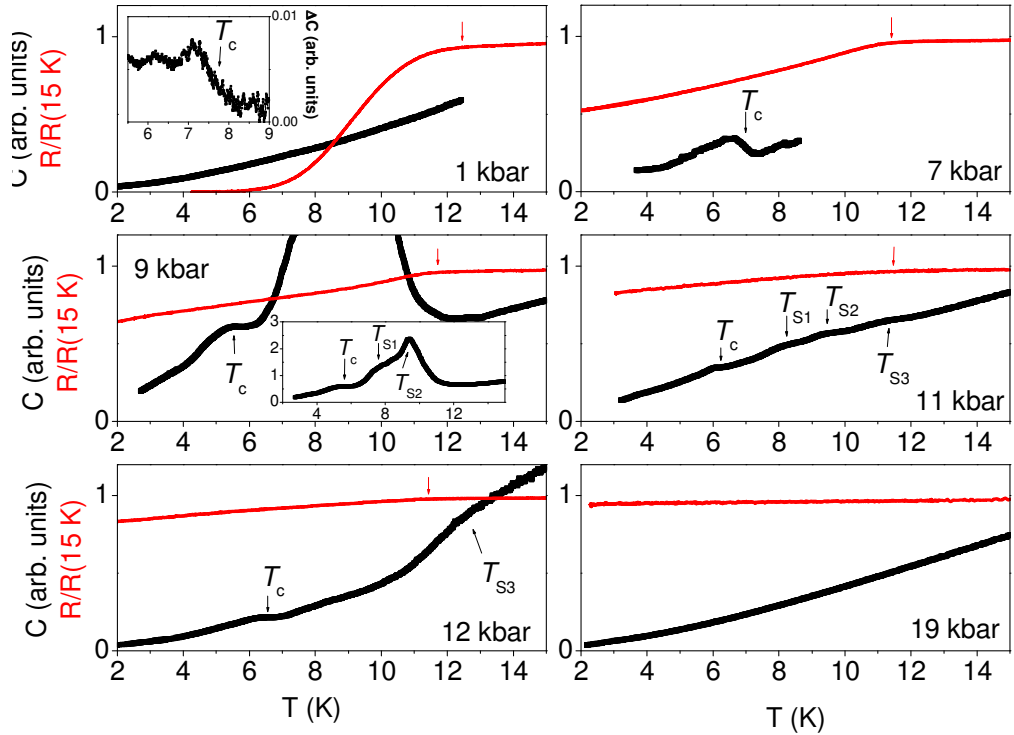

**Figure 2b)** Specific-heat (black) in arbitrary units and normalized electrical resistance (red) for selected applied pressures. The black arrows indicate the superconducting transition ($T_c$) in the specific heat, $T_{S1} - T_{S3}$ indicate additional anomalies related to structural phase transitions. The red arrows mark the onset of the resistive transition. The insets show details of the specific heat on a different scale. The data at 1 kbar was measured just after closing the cell. For technical reasons all following measurements in higher pressure have been done upon releasing the pressure starting from 24 kbar.

Figure 2 shows the electrical resistance and the specific heat of Ca122 at various applied pressures. The resistance data has been normalized by its value at 15 K. The variation of the 15 K resistivity with pressure is included in absolute values in Figure 3. A broad drop to zero resistance is seen for 1 kbar pressure below 12 K, which was measured just after closing the pressure cell. For technical reasons we applied in the following a pressure of 24 kbar and measured all further data upon gradually releasing the pressure. A very weak downturn of the resistance starts to become visible in 19 kbar, which gets more and more pronounced in lower pressure. Apart from the data in the low initial pressure of 1 kbar, no zero resistance is reached down to the lowest accessible temperature of 2 K. The onset of the downturn varies only between 11 K (highest pressure) and 12 K (lowest pressure). The dropping resistance below 11 - 12 K certainly indicates the presence of superconducting correlations in the pressure range between 1 and ~20 kbar and show that an (at least percolative) path of zero resistivity is found in 1 kbar.

The absence of zero resistivity in higher pressure has been observed before, although in our case this occurs already at comparatively low pressure. This could be related to the loss of a percolative path related to structural phase separation. However, as the specific heat anomaly indicates that a major part of the sample's volume is superconducting in this pressure range, this is unlikely. Most of our data has been recorded upon decreasing pressure. We have observed a similar behavior previously in cuprate superconductors, where zero resistivity was lost upon releasing the pressure whereas the specific heat still indicated a bulk superconducting state. Therefore we attribute this behavior rather to micro-cracks in the sample which cause dissipation in the superconducting state. The presence of micro-cracks is however not expected to influence the specific-heat.

Resistivity does not provide information on the establishment of bulk superconductivity. In the specific heat in 1 kbar a small broad jump centered around 7.7 K is found. Its magnitude represents only 3 % of the total specific heat. For clarity we subtracted a phonon background in the insert of Fig 2. The phonon background estimation was extracted from data taken in 24 kbar, where the sample is in the normal metallic state at all temperatures. The smallness of the jump at this pressure may indicate that only a fraction of ~5-10 % of the sample's volume is superconducting. Although its magnitude is small, a visible anomaly in this bulk thermodynamic quantity cannot be attributed to a filamentary type of superconductivity, which would be associated with a negligible volume fraction of the sample only. Between 1 and 7 kbar the jump sharpens and grows to a size of ~1/3 of the total specific-heat magnitude ($T_c$ = 6.9 K). As the specific heat probes the entire volume of the sample, this clearly indicates a bulk superconducting state. The temperature range is limited here to 8.7 K as this represented the final measurement of the series during which an electrical contact was lost.

In 9 kbar several anomalies appear in the specific heat. A large peak appears at 9.4 K with a pronounced shoulder around 7.6 K. A similar step-like anomaly as in the lower pressure is centered around 5.7 K. The onset of the peak-like anomaly falls together with the downturn of the resistivity at 12 K and the anomaly extends over the temperature range where the resistivity drops. From the comparison with the resistivity it is clear that the entire anomaly is somehow related to the superconducting transition. However, the compared to the phonon background large magnitude of the anomaly and the peak-like first-order nature of the transition indicate that its origin is not solely from superconducting degrees of freedom. At the same pressure a pronounced decrease in the residual resistivity (see Figure 3) is observed, hence it is most likely that a structural transition approaches the superconducting state at this pressure causing the peak–like anomaly. The true superconducting specific-heat transition is rather the step-like second order anomaly at 5.7 K which resembles a BCS–like superconducting transition. The structural transition may be associated with strain which causes superconducting precursors above the true bulk superconducting transition temperature. A similar peak-like anomaly has been observed in $BaFe_2As_2$ in the specific heat under a pressure for which it is known that a structural / magnetic transition meets the superconducting transition line [35]. In the pressure range between 3 and 5 kbar several structural phase transition lines merge and may approach zero temperature in this pressure range [36]. In 11 kbar the magnitude of the anomaly decreases and splits in 3 distinct bumps at 8, 19 and 12 K. The BCS-like $T_c$ anomaly is still found at 6 K but the size of the jump is smaller than at lower pressures, which indicates that superconductivity is weakened in this pressure range. In only slightly higher pressure (12 kbar) a further broad upward step-like anomaly occurs at 12 K. At higher pressure no more signs of a superconducting transition and no further anomalies are observed. In Figure 3 we summarize the data in form of a low-temperature phase diagram of Ca122 under pressure.

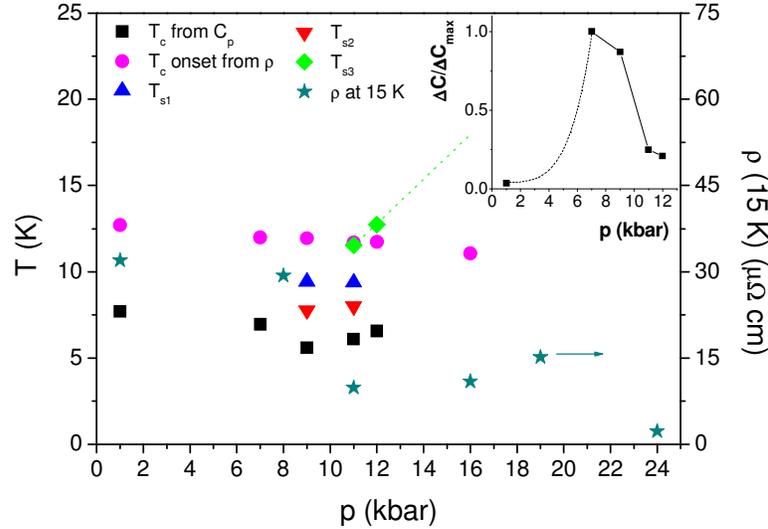

**Figure 3.** Temperature vs. pressure phase diagram of $CaFe_2As_2$ derived from our data. Squares: bulk superconducting transition temperature $T_c$ as obtained from the specific heat. Open circles: Onset of the resistively determined superconducting transition. The triangles and diamonds represent specific-heat anomalies of structural phase transitions. Stars: 15 K resistivity values (plotted against the right-hand axis). Inset: normalized jump size of the superconducting anomaly in the specific heat which is closely related to the superconducting condensation energy.

Superconducting correlations occur below 12 K over a rather large pressure range from 1 to 16 kbar, which is consistent with phase diagrams in literature based on resistivity data [19,26]. A small specific-heat anomaly at $T_c$ is already observed in 1 kbar pressure and a large anomaly is found between 7 and 12 kbar. The size of the specific-heat anomaly $\Delta C$ (see inset of Figure 3) is closely related to the superconducting condensation energy [28,37] and a large anomaly indicates that the majority of the volume of the sample is in the superconducting state (compared to other 122 compounds, which superconduct at ambient pressure, we estimate that at least 50 % of the volume becomes superconducting).

In most of the proposed phase diagrams in literature it appears as if the two structural transition lines related to the transition from the T to the O phase and the T to cT phase do not approach low temperatures [15,26]. The pronounced additional anomalies in the specific heat occur however in the pressure range where these two transition lines meet and indeed a pressure-induced transition from the O to the cT phase must occur at low temperatures. This is confirmed by Ref. [36]: the transition line into the high-pressure cT phase drops to zero temperature at ~ 4 kbar pressure upon increasing pressure and separates the magnetic low-pressure O phase from the high-pressure cT phase. Most likely, the anomalies we observe are related to this transition line. This pressure-induced structural transition causes a significant drop of the resistivity at 15 K towards higher pressure at a pressure of ~5 kbar, which is also present in our data. A dramatic increase in the size of the superconducting specific-heat anomaly occurs close to 7 kbar, which is certainly related to this pressure-induced structural phase change. This is further confirmed by the fact that the resistivity starts to decrease at the onset of this structural specific-heat anomaly. The absence of the cT transition upon releasing pressure [36] has been discussed with the possibility of having a strain-induced mixed phase of the O and cT phase located around this transition line. We cannot provide a clear statement from our data but the fact that the structural transition anomaly splits into 3 distinct anomalies at 11 kbar may indeed be a sign that different crystallographic phases coexist within a certain pressure range.

To summarize, our specific-heat data measured in a Bridgman-type of pressure cell allow us, based on a true bulk thermodynamic quantity, to definitely conclude that pressure-induced superconductivity in Ca122 is certainly not a filamentary superconductivity originating from structural domain boundaries but a bulk property. Although we cannot rule out that a phase separation of different coexisting crystallographic phases may exist, we can estimate that at least 50 % of the volume becomes superconducting. The bulk superconducting transition remains at temperatures below 7 K, whereas the onset of the broad resistive transition remains robustly at 12 K. This demonstrates the necessity of using a bulk thermodynamic method for the exact determination of superconducting transition temperatures of iron pnictides superconductors, especially under the influence of pressure. A pronounced structural anomaly occurs in the pressure range from ~7 kbar to 12 kbar, which we attribute to a pressure-induced transition from the low-pressure O phase into the high-pressure cT phase, which is approaching the superconducting phase. It runs over a certain pressure range (9 and 12 kbar) parallel to the superconducting transition line before it levels off towards higher temperature. The onset of this structural transition falls together with the onset of the resistive superconducting transition, indicating that strain-induced superconducting precursors may exist well above the bulk specific-heat transition and cause the strong discrepancy between the resistive and thermodynamic determination of the superconducting transition temperature.

**Acknowledgments:**
We thank U. Lampe for technical support. This work was supported by the Research Grants Council of Hong Kong Grants DAG_S08/09.SC05 and 603010.